\documentclass[preprint,amsmath,amssymb,showpacs,pra]{revtex4-1}

\usepackage{graphicx}
\usepackage{mathrsfs}
\usepackage[colorlinks]{hyperref}
\usepackage{dcolumn}
\usepackage{bm}
\usepackage{setspace}

\begin{document}
\title{Dynamical excitations in the collision of two-dimensional Bose-Einstein condensates}
\author{T. Yang}
\author{B. Xiong}
\author{Keith A. Benedict}
\affiliation{Theory of Condensed Matter, School of Physics and Astronomy, University of
Nottingham, Nottingham NG7 2RD, United Kingdom}
\date{\today}

\begin{abstract}
We investigate the way in which the pattern of fringes in a coherent pair of two-dimensional Bose condensed clouds of ultra-cold atoms traveling in opposite directions subject to a harmonic trapping potential can seed the irreversible formation of internal excitations in the clouds, notably solitons and vortices. We identify under, over and critically damped regimes in the dipole oscillations of the condensates according to the balance of internal and centre-of-mass energies of the clouds. We carry out simulations of the collision of two clouds with respect to different initial phase differences in these regimes to investigate the creation of internal excitations. We distinguish the behaviour of this system from previous studies of quasi one-dimensional BEC's. In particular we note that the nature of the internal excitations is only essentially sensitive to an initial phase difference between the clouds in the overdamped regime.
\end{abstract}

\pacs{03.75.Kk, 03.75.Lm}

\maketitle

\section{Introduction}
Andrews et al \cite{Science.275.637} clearly demonstrated the matter-wave nature of atomic Bose condensates by showing that when two Bose condensed clouds are allowed to overlap spatially, quantum interference fringes are observed. This, and most subsequent experiments on atomic interference, used freely expanding atom clouds \cite{interf,PRL.81.5477,PRL.87.080402,PRA.64.031601} to avoid the complicating effects of interactions. An alternative approach, however, involves controlling the motion of atom clouds using time-dependent trapping potentials \cite{PRA.64.063607,PRL.92.050405}, often generated by atom chips \cite{RMP.79.235}, where the nonlinear effect due to interatomic interactions would be significant. In the noninteracting limiting case the interference fringe pattern will appear while the clouds are spatially overlapping but smoothly disappear as the clouds separate. However, in Ref.\cite{PRA.68.013611} it was shown that the recombination process of a split condensate is very sensitive to atomic interactions. When interactions are important, the spatial modulation associated with the interference pattern can lead to the creation of non-trivial internal excitations (solitons and vortices) which do not disappear after the clouds separate. These excitations may have important consequences for interferometry. Such processes have been widely discussed in the context of quasi one-dimensional (1D) BEC clouds \cite{PRL.83.5198,PRL.87.080402,JPB.37.L385,PRL.98.180401,PRL.100.100402,PRA.79.063624}. The relative phase of two BECs can be read out by measuring the size of the cloud \cite{PRL.98.180401,NatPhys.1.57} after free expansion or the amplitude of soliton oscillations (or the condensate dipole mode) \cite{JPB.37.L385}. Analysis of the dynamics of solitons highlighted the potential for exploiting the resonant production of vortices when the initial relative phase of two interfering condensates is $\pi$ to develop motion detectors \cite{PRL.100.100402}. In a two-dimensional (2D) configuration (disk-shaped condensates) the situation may be different. Such a 2D geometry is of interest because, like the elongated (quasi-1D) geometry studied previously, 2D condensates have large length (in both $x$- and $y$-directions for 2D configurations): a desirable feature for matter wave interferometry because it provides natural averaging over any small fluctuations in the trap potentials, the force to be measured, or the thermal fluctuations within the BEC. However, in 2D one expects the effects of (quantum and thermal) phase fluctuations to be less severe.

The primary objective of this paper is to investigate the processes by which internal excitations are seeded by the formation of quantum interference fringes in a coherent pair of 2D Bose condensed clouds in relative motion subject to a harmonic trapping potential and how these result in the damping of dipole oscillations: in particular, the sensitivity of the dynamics to the phase difference between the two clouds. In addition to the intrinsic interest of such processes, they are significant (and mostly inimical) to applications of matter-wave interferometry such as metrology, gravitational, temporal and rotational sensing \cite{tlgus,sgupta,yjwang} as well as fundamental studies of decoherence \cite{hugbart} and vortex production \cite{PRL.83.2498,PRL.87.080402,PRL.81.5477,PRA.64.031601}.

The energy cost of the formation of internal excitations must be paid from the centre-of-mass (c.m.) kinetic energy of the clouds hence leading to the damping of c.m. oscillations in a confining potential. The dynamics of the system is mainly determined by the competition between the c.m. kinetic energy and the interatomic interaction energy.  The system is studied in three regimes. (a) The underdamped regime corresponds to the situation in which the c.m. kinetic energy is much larger than the interatomic interaction energy. In this case the system responds in a manner that is qualitatively similar to the non-interacting system, albeit with curved fringes due to the non-uniformity of the clouds. (b) The critically damped regime refers to the situation in which the c.m. kinetic energy and the interatomic interaction energy are of the same order. This means that nonlinearity is important and the effect of scattering becomes noticeable which may make the dynamics of the system complicated. (c) In the overdamped regime, the c.m. energy is much smaller than the interaction energy, where the system undergoes a merging process dominated by the nonlinear interaction rather than collision.

In this paper, we present detailed calculations within the framework of zero-temperature mean-field theory, which enable us accurately to characterize the behaviour of collision/interference dynamics as a function of the initial displacement between the two disk-shaped condensate clouds and the role of the initial relative phase difference in these different regimes. We will describe the way in which quantum interference can lead to the creation of permanent excitations and the consequent damping of dipole oscillations.

The physical situation that we consider in this paper involves cooling a cloud of atoms to zero temperature in an anisotropic harmonic potential. The potential is then changed to have two minima in order to (partially or fully) separate the condensate into two clouds whose centres of mass are symmetrically placed about the minimum of the original potential, a distance $2\Delta$ apart. It is supposed that the separation process is carried out adiabatically so that the clouds retain their mutual phase-coherence and remain in the ground state of the double well potential. We allow for the possibility that an overall phase difference, $\theta$, is imposed between the two clouds. The potential is then suddenly switched back to its original form and the system is allowed to evolve. The stipulation that the two clouds retain their full phase coherence ensures that the whole system is described by a single, complex, order parameter field which evolves according to the time-dependent Gross-Pitaevskii (GP) equation\cite{gp1,gp2}.

A Quasi-2D BEC can be formed in a highly anisotropic trap of the form
\begin{equation}
\begin{split}
V_{\text{trap}}\left(\textbf{r}\right) & = V_{\perp}(\textbf{r}_\perp)+V_z(z)\\
& = \frac{1}{2} m \omega_{\perp}^2 \left(x^2 + y^2\right)
+ \frac{1}{2} m \omega_z^2 z^2\qquad,
\end{split}
\end{equation}
with $\omega_z \gg \omega_{\perp}$. In this limit the dependence on the $z$-coordinate is quenched such that the 3D order parameter for a gas of $N_{\text{tot}}$ atoms has the form
\begin{equation}
\Psi(x,y,z;t) = \sqrt{N_{\text{tot}}}\phi(x,y;t)\exp\{-i \omega_z t / 2 \} \exp\{ - z^2 / 2 l_z^2\}/(\pi l_z^2)^{1/4}
\end{equation}
where $l_z=\sqrt{\hbar/m\omega_z}$ is the oscillator length in the tightly confined direction and $\phi$ is normalized to unity.

The dynamics of such a 2D BEC with $N_{\text{tot}}$ atoms is governed by the 2D GP equation
\begin{equation}\label{GPE-2D}
i\hbar\frac{\partial}{\partial t}\phi=-\frac{\hbar^2}{2m}\nabla_{\perp}^2\phi+V_\perp(\bf r_\perp) \phi+g_{2}N_{\text{tot}}\left\vert\phi\right\vert^2\phi\qquad,
\end{equation}
where $g_{2}=g_{3D}/\sqrt{2\pi}l_z=2\sqrt{2\pi}a_s l_z\hbar\omega_z$ is the 2D coupling constant and $a_s$ is the bulk $s$-wave scattering length.

Initially, we suppose that the system is prepared as described above and that the two clouds are widely separated so that they do not overlap spatially. At time $t=0$ we have
\begin{equation}
\phi(x,y;t=0) = \frac{1}{\sqrt{2}}\Bigl(\phi_0\left(x+\Delta,y\right) + e^{i \theta}\phi_0\left(x-\Delta,y\right)\Bigr) \label{initial}
\end{equation}
where $\phi_0(x,y)$ is a localized wave-packet.
This is a coherent superposition of two independent, normalized, condensate wave-functions $\phi_0(x\pm\Delta,y)$.
In the next two sub-sections we will discuss the subsequent evolution of the order parameter, first without and then with interactions. We will leave the discussion of the situation in which there is an initial overlap between the two clouds to later in the paper.

\subsection{Non-Interacting Case}
If a non-interacting BEC of $N/2$ atoms is prepared in a displaced trap with $V_{\perp}\left(x, y\right)=m\omega_{\perp}^2\left(\left(x-\Delta\right)^2 + y^2\right)/2$  by cooling to its ground state and the trap is then suddenly switched so that the minimum moves to $x=y=0$ then the cloud c.m. will execute undamped simple harmonic motion. The initial state, in this case, is a simple gaussian, centred on the point $x=\Delta$, $y=0$ which evolves according to
\begin{equation}
 \phi(x,y;t) = e^{-i\chi(t)}\phi_0\left(x-X(t),y\right) e^{iP(t)x/\hbar}\label{eq_mo}
\end{equation}
where
\begin{equation}
\begin{split}
X(t) & = \Delta \cos \left( \omega_{\perp} t \right)\qquad,\\
P(t) & = -m \omega_{\perp} \Delta \sin \left( \omega_{\perp} t \right)\qquad,
\end{split}
\end{equation}
\begin{equation}
\phi_0(x,y) = \sqrt{\frac{1}{\pi l^2_{\perp}}}\exp\Bigl\{-\left(x^2+y^2\right)/2l_{\perp}^2\Bigr\}\qquad,\label{gaussian}
\end{equation}
with $l_{\perp}^2=\hbar/m\omega_{\perp}$ and $\chi(t)$ is a time dependent phase. $X(t)$ and $P(t)$ are, of course, the position and momentum of a classical particle of mass $m$ initially at rest at position $x=\Delta$ in the same potential.

If the initial state consists of a superposition of two identically prepared gaussians (Eq. (\ref{gaussian})) displaced by $x\mapsto x\pm\Delta$ then, in the absence of interactions, they will evolve independently as
\begin{equation}
\phi(x,y;t) = \frac{1}{\sqrt{2}} \left(e^{-iP(t)x/\hbar} \phi_0(x+X(t),y) + e^{i\theta} e^{iP(t)x/\hbar} \phi_0(x-X(t),y) \right)
\end{equation}
with $X(t)$ and $P(t)$ as given above. The corresponding density, $\rho_2(x,y;t)=N\left\vert\phi(x,y;t)\right\vert^2$ is then
\begin{equation}\begin{split}
\rho_2(x,y;t) & =\frac{N}{\pi l_{\perp}^2} e^{ - \left( x^2 + y^2 + X^2(t) \right) / l_{\perp}^2 } \left\{ \cosh \left( \frac{2X(t) x }{ l_{\perp}^2} \right) + \cos \left( \frac{ 2 P(t) x}{\hbar} + \theta \right) \right\} \qquad.
\end{split}\end{equation}
The two clouds are maximally overlapped at times $T_n=\left(n+\frac{1}{2}\right)\pi/\omega_{\perp}$ when
\begin{equation}
\rho_2(x,y;T_n) = N\frac{1}{\pi l_{\perp}^2}e^{-\left(x^2+y^2\right)/l_{\perp}^2}\left\{1 + \cos\left(\frac{2m\omega_{\perp}\Delta}{\hbar}x - \theta\right)\right\}
\end{equation}
corresponding to perfect rectilinear fringes parallel to the $y$-axis with the fringe spacing given by the de Broglie relation $L={2\pi\hbar}/{2m\omega_{\perp}\Delta}$. The initial relative phase difference simply shifts the fringe pattern.

If we consider the energetics of the non-interacting case we can express the energy of the initial state as $E_T = E_{K} + E_{P}$ where
\begin{equation}
E_K = N \int d^2{\bf r_\perp} \frac{\hbar^2}{2m} \left\vert \nabla_{\perp} \phi \right \vert^2 = N \frac{1}{2m} \left( P(t) \right)^2 + N \frac{1}{2} \hbar \omega_{\perp}
\end{equation}
is the kinetic energy and
\begin{equation}
E_P = N \int d^2{\bf r_\perp} \frac{1}{2} m \omega_{\perp}^2 \left\vert {\bf r_\perp} \right\vert^2 \left\vert \phi \right\vert^2 = N \frac{1}{2} m \omega_{\perp}^2 \left( X(t) \right)^2 + N \frac{1}{2} \hbar \omega_{\perp}
\end{equation}
the (trap) potential energy. We can regroup these into two conserved quantities
\begin{equation}
 E_{cm} = N\frac{1}{2}m\omega_{\perp}^2\Delta^2
\end{equation}
is the energy associated with the bulk motion of the two clouds and
\begin{equation}
 E_0 = N\hbar\omega_{\perp} = N\frac{\hbar^2}{m l_{\perp}^2}
\end{equation}
is the zero-point energy of the two clouds.

\subsection{Interactions and Damping}
The inclusion of interactions between the atoms will, naturally modify this picture. While a single BEC, when subject to a displaced trap potential, will execute undamped simple-harmonic motion (dipole mode), the complexity of the dynamics of a pair of separated condensates will depend on the relative magnitudes of the c.m. energy and the interatomic interactions. Interatomic interactions will lead to the irreversible creation of internal excitations within the clouds and hence damping of the c.m. oscillations. We therefore identify underdamped, critically damped and overdamped regimes. In the underdamped regime, the clouds retain their separate identity, separating after a clear ``collision'' process and then recolliding a number of times with decaying amplitude. In the overdamped regime the oscillation is effectively stopped by the first collision leaving a single, merged condensate cloud. In the critically damped regime the clouds pass through one another but never properly separate after the first collision.

If we imagine adiabatically splitting an interacting 2D BEC with $N$ atoms into two well separated clouds, then for moderate interaction strengths we would expect the initial clouds to be well-described by the Thomas-Fermi (TF) approximation appropriate for the minima of the preparation trap potential
\begin{equation}
 |\phi_0(\bf r_\perp)|^2 = \left\{\begin{array}{cc}
\frac{m\omega_{\perp}^2}{g_{2}N}\left(R_{TF}^2 - |\bf r_\perp|^2\right) & \left\vert{\bf r_\perp}\right\vert<R_{TF}\\
0 & \left\vert{\bf r_\perp}\right\vert > R_{TF}
\end{array}\right.
\end{equation}
where, as always, the TF radius, $R_{TF}$, is chosen to give the correct total number of particles which yields
\begin{equation}
 R_{TF} = \sqrt{2}l_{\perp}\gamma
\end{equation}
(which must satisfy $R_{TF}\ll\Delta$) and the 2D TF parameter is
\begin{equation}
\gamma = \left(\frac{2N a_s}{l_z \sqrt{2\pi}}\right)^{1/4}\qquad.
\end{equation}
In this case, the conserved energy will be $E_T = E_{\text{cm}} + E_{\text{int}}$ where
\begin{equation}
\begin{split}
E_{\text{cm}} & = N\frac{1}{2}m\omega_{\perp}^2\Delta^2\\
E_{\text{int}} & = E_{TF} = \frac{2}{3}\mu_{TF}N = \frac{2}{3}N\hbar\omega_{\perp}\gamma^2
\end{split}
\end{equation}
These will not be separately conserved and we expect an irreversible transfer of energy from $E_{cm}$ to $E_{\text{int}}$ leading to damping of the dipole oscillations of the two condensates at the expense of the creation of internal excitations within each cloud.

We seek to define an effective measure of the relative importance of interactions which can be directly evaluated in a simulation. We define a parameter, $\eta_n$, at each turning point $T_n=n\pi/\omega_{\perp}$ of the dipole oscillation (at which the c.m. kinetic energy is zero) as
\begin{equation}
\eta_n = \frac{E_{I}}{E_P} = \frac{ \frac{1}{2} g_{2}N^2 \int \left\vert \phi( {\bf r_\perp}, T_n ) \right\vert^4 d^2 {\bf r_\perp} }{ \frac{1}{2}m \omega_{\perp}^2 N  \int  \left\vert {\bf r_\perp}\right\vert^2\left\vert\phi({\bf r_\perp},T_n)\right\vert^2d^2{\bf r_\perp}}\qquad.
\end{equation}
In the non-interacting limit ($g_{2}=0$) this is identically zero. In the TF regime it will be
\begin{equation}
\eta_n \sim \frac{\frac{2}{3}\hbar\omega_{\perp}\gamma^2 + \epsilon_{\text{EX}}}{\frac{1}{2}m\omega_{\perp}^2\Delta_n^2}
\end{equation}
where $\Delta_n$ is the displacement of the c.m. of the cloud from the centre of the trap at time $T_n$ and $\epsilon_{\text{EX}}$ is the excitation energy per atom of the cloud (i.e. the excess internal energy of the cloud over the TF energy per particle). Since the scattering will act to damp the dipole oscillations (decreasing $\Delta_n$) and create excitations within the cloud (increasing $\epsilon_{\text{EX}}$) $\eta_n$ will increase with $n$.

\section{Preparation of Initial State}
In experiments, a condensate held in a magneto-optical trap (MOT) can be split into two condensates adiabatically by using a tailored magnetic potential from an atom chip \cite{sy}, by the introduction of an optical barrier via a shaped blue-detuned laser \cite{scherer} or by passing counterpropagating red-detuned laser beams through an acousto-optic modulator driven at RF frequency \cite{PRL.98.180401} to form a double well potential in a plane. If the plane is not horizontal, the difference in gravitational potential between the two local minima leads to a time dependent phase difference between the two condensates\cite{PRL.98.180401}. Alternatively, a phase difference can be imposed between the clouds by phase imprinting\cite{PRA.60.R3381}.

We based our simulations on a condensate of $N=2\times10^4$ ${}^{87}R_b$ atoms in a trap potential with $\omega_{\perp}=2\pi\times5H_z$ and $\omega_z = 2\pi\times100H_z$. We will always keep the interaction strength and the number of atoms in the system fixed. The only quantities that we change are the initial displacement, $\Delta$ and the initial relative phase of the condensate clouds, $\theta$. We use a scattering length, $a_s=0.3\times5.4nm$, which is shorter than that appropriate for $ ^{87}Rb$, in order to see the structure of vortex excitations clearly in our simulation. To consider the effect of the relative phase on the dynamics, we keep the initial phase of one condensate fixed and adjust the initial phase of the other from 0 to $2\pi$.

To simulate the underdamped and critically damped cases, we obtained the configuration, $\phi_0(x,y)$, of the order parameter which minimized the mean field energy
\begin{equation}
\int \left\{\frac{\hbar^2}{2m}\left\vert \nabla_\perp\phi_0\right\vert^2 + \frac{1}{2}m\omega_{\perp}^2\left\vert \bf r_\perp\right\vert^2 \left\vert \phi_0\right\vert^2 +\frac{N}{2}g_{2}\left\vert\phi_0\right\vert^4\right\} d^2 \bf r_\perp
\end{equation}
subject to the constraint $\int d^2{\bf r_\perp} \left\vert\phi_0\right\vert^2 = 1$. The initial state for the simulations was that given in Eq. (\ref{initial}). We used this method to generate the initial state rather than minimizing the mean-field energy of the system in a suitable quartic potential in order to improve numerical efficiency and accuracy. The order parameter was then allowed to evolve in the 2D trap potential.
\begin{figure}[t]
  \centering
   \includegraphics[scale=0.8,bb=-20 350 779 580, clip]{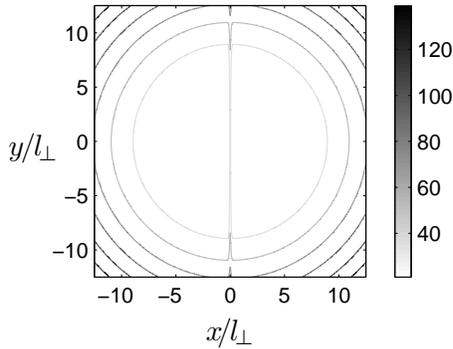}
   \caption{The equipotential lines of a 2D double well potential with narrow cental barrier along about $x=0$. The parameters used are $\sigma_x/l_{\perp}=0.1$, $\sigma_y/l_{\perp}=10$, $U_0/\hbar\omega_{\perp}=50$.  }\label{dw-pot}
\end{figure}

To reach the overdamped regime without changing the frequency of the harmonic trap required $ \Delta < R_{TF} $ so another method of preparing the initial state was used. The lowest energy configuration of the order parameter was found in a potential
\begin{equation}
V_{DW}(x, y) = \frac{1}{2} m \omega_{\perp}^2 (x^2+y^2) + U_0 \exp\left\{-\frac{x^2}{\sigma_x^2} - \frac{y^2}{\sigma_y^2}\right\}.\label{DW}
\end{equation}
where the second term provides a strong central barrier to suppress the overlap between the two clouds. Such a trap potential would be a suitable model of, for example, the effect of a narrowly focused, blue-detuned laser applied to split the condensate. The height of the central barrier is given by the amplitude of the Gaussian function $U_0$, and the distance between the two wells can be adjusted using the parameters $\sigma_x$ and $\sigma_y$. We can see two distinct potential minima clearly in Fig. \ref{dw-pot}, which have the same shape as the RF-induced potential in the experiment in Ref.\cite{schumm}. The parameters we used in our simulations are $\sigma_x/l_{\perp}=0.1$, $\sigma_y/l_{\perp}=10$, and $U_0/\hbar\omega_{\perp}=50$. The ground state order parameter of the condensate is obtained by the imaginary-time-evolution method \cite{chiofalo} for the time-dependent GP equation (\ref{GPE-2D}). We then used this configuration as the initial state for real-time evolution of the GP equation, representing the sudden removal of the laser. The drawback of this method is that the change of potential on its own will excite shape oscillations in the cloud. The alternative method would have been to prepare the clouds in a higher frequency trap
potential and then allow the system to evolve in a softer trap. Even in the absence of interactions this would have led to the clouds expanding and changing shape during the slow approach to the collision.

\begin{figure}[t]
  \centering
   \includegraphics[scale=0.6,bb=-60 300 779 580, clip]{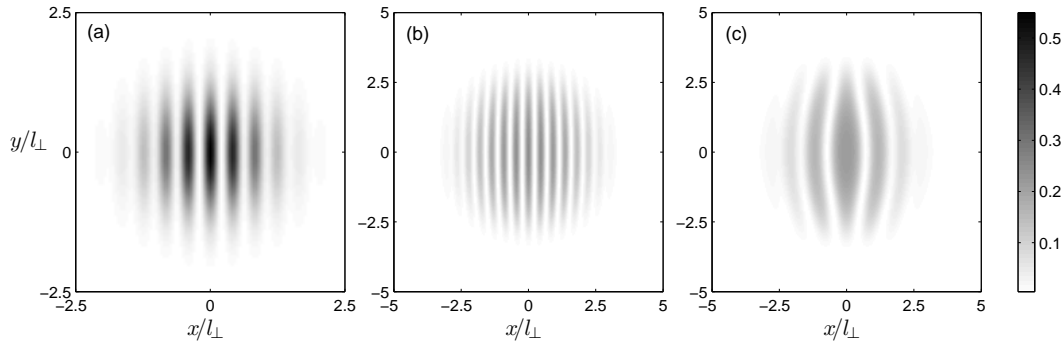}
   \caption{Normalized density profiles of the two condensate clouds without initial phase difference colliding at the bottom of the trap. (a) noninteracting case with $\Delta = 7.5l_{\perp}$; (b) interacting case with $\Delta = 7.5 l_{\perp}$; (c) interacting case with $\Delta = 4 l_{\perp}$. }\label{fringe}
\end{figure}

\section{Numerical solution and results}
In the numerical investigations we used $l_{\perp}$ as the unit of length and $1/\omega_{\perp}$ as the unit of time. Hence we define dimensionless wavevectors $\tilde{k}=kl_{\perp}$, times $\tilde{t} = \omega_{\perp}t$ and frequencies $\tilde{\omega}=\omega/\omega_{\perp}$
in order to minimize numerical instabilities.

\subsection{Scattering effects in the underdamped regime}
As we have discussed, the importance of the scattering between atoms depends on the competition between the c.m. kinetic energy and the interatomic interaction energy, which in turn affect the dynamics of the system.

In Fig. \ref{fringe}, we show the density profiles at the point of maximum overlap ($\tilde t=\pi/2$) of the two
clouds with zero initial phase difference starting from different displacement positions. The bright region is a
low density area, while the dark region is a high density area. For a noninteracting system there is only kinetic
energy so $\eta=0$. In this linear regime the unperturbed condensates always show straight fringes during the
interference as seen in Fig. \ref{fringe}(a). For interacting systems when the displacement is large, the clouds will
have maximum c.m. kinetic energy, i.e. high relative peak velocity when they begin to interfere. The system still
responds linearly. The scattering processes arising from interactions produce a frictional force on the cloud tending to reduce its velocity and hence increase the fringe spacing. The frictional force will be largest in the centre of the trap where the density is highest. In addition, the interaction will lead to an over-pressure in the fringe maxima which will tend to fatten the fringes in the trap centre. However the duration of the collision is correspondingly short, then the effect of scattering is very small. In Fig. \ref{fringe}(b), the initial separation is set to be $\Delta = 7.5 l_{\perp}$. We can see that interference fringes are nearly straight. With smaller initial separations, the two clouds reach a smaller peak velocity, the interaction energy is comparable to the c.m. kinetic energy and the duration of the collision is correspondingly longer. As is shown in Fig. \ref{fringe}(c), where $\Delta = 4 l_{\perp}$, the distortion of the fringes has become quite sizable. The width of the fringes increases with the nonlinearity, while the peak density does not change dramatically in comparison to that of the fringes in the linear situation (Fig. \ref{fringe}(b)). The nonlinear situation however gives us fewer fringes associated with the damping of the dipole oscillation of each cloud. The expansion rate of the central fringe is highly nonuniform along $y$-axis due to the non-uniform density distribution. At the centre of the clouds, where the density is highest, the fringes expand most. The spacing of fringes are close to their non-interacting value towards the edge of the cloud (low density region) where the interaction is negligible. This leads to the central fringe taking a lenticular shape, accompanied by the bending of the other fringes. There is only the central fringe left eventually before the two clouds start to separate.

\begin{figure}[t]
  \centering
   \includegraphics[scale=0.7,bb=-40 311 779 600]{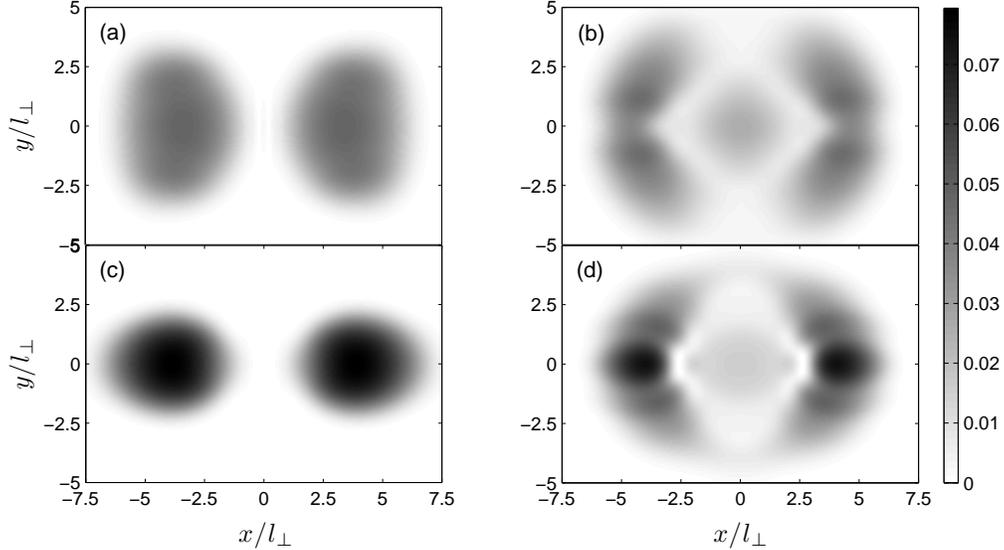}
   \caption{Normalized density profiles of the colliding condensates in the trap without initial phase difference. The initial separation $\Delta$ is $7.5 l_{\perp}$ (a)(c) and $4 l_{\perp}$ (b)(d) respectively. Times are (a) $\tilde t = 2.15$, (b) $\tilde t = 3.05$ and (c)$\tilde t = 4.25$, (d) $\tilde t = 3.5$. }\label{fringe1}
\end{figure}

In Fig. \ref{fringe1} we show the density profiles of the two clouds without initial phase difference after their first collision, where the initial separation $\Delta$ is $7.5 l_{\perp}$ (Fig. \ref{fringe1}(a) and (c)) and $4 l_{\perp}$ (Fig. \ref{fringe1}(b) and (d)) respectively. The effect of interactions is not obvious in the underdamped regime ($\Delta=7.5l_{\perp}$). As seen in Fig. \ref{fringe1}(a) the condensates expand a little bit in the $y$-direction which means that some atoms gain a $y$-component to their momentum via scattering. This momentum is small, however, due to the short collision time, which means that the scattering induced distortion of condensates is not obvious after the first collision. After the first collision the clouds pass through each other; each moving towards the original position of the other. The density profile of the condensates will vary in a manner similar to the breathing mode and the condensates will undergo a damped c.m. oscillation. Fig. \ref{fringe1}(c) shows that the clouds regain their original shape with a small distortion at $\tilde t=4.25$ before their second collision.

\begin{figure}[t]
  \centering
   \includegraphics[scale=0.7,bb=-80 280 779 600,clip]{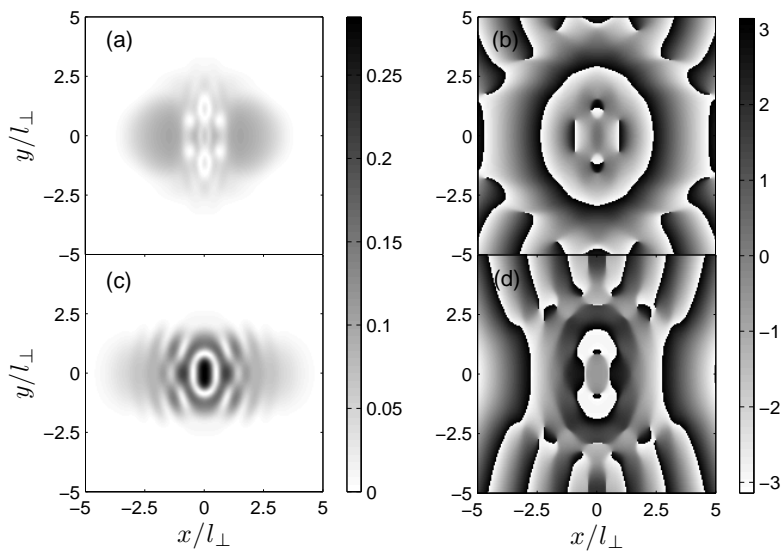}
\caption{Normalized density profiles (the left panels) and the corresponding phase plots (the right panels) of the colliding condensate clouds in a trap during the second recombination. This is the continuation of the simulation for $\Delta=4l_\perp$ shown in Fig. \ref{fringe1}. From (a) and (b) we can see clearly the formation of two vortex pairs located at the left and the right side of the centre of the trap. Curved solitons arise in the centre of the trap. These vortex pairs approach each other until the vortices with opposite charges annihilate, leaving a soliton ring as seen in (c) and (d). The soliton ring will decay into vortices thereafter. Times are (a,b) $\tilde t=4.15$, (c,d) $\tilde t=4.5$.}\label{exc-lv}
\end{figure}

\begin{figure}[t]
  \centering
   \includegraphics[scale=0.6,bb=-80 250 779 600,clip]{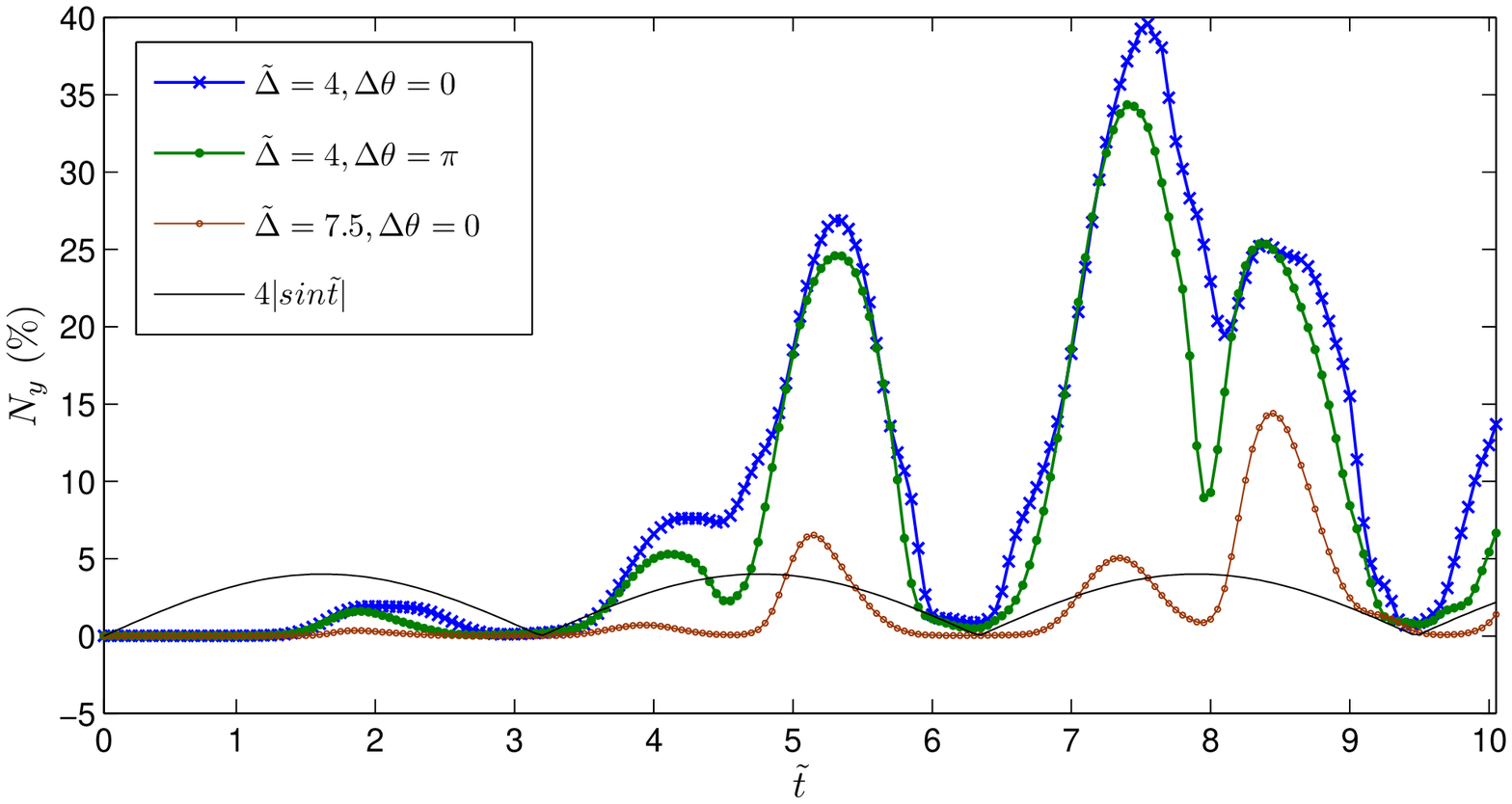}
   \caption{(Color online) Oscillations of condensate atom number $N_y$ for different initial conditions. }\label{collision}
\end{figure}

When the initial separation is reduced, the maximum c.m. kinetic energy is correspondingly smaller and the nonlinear
interaction term in Eq. (\ref{GPE-2D}) becomes more important. We reach the critically damped regime
(e.g., $\Delta=4l_{\perp}$) where the cloud c.m. kinetic energy is comparable to the interaction energy. As seen in
Fig. \ref{fringe1}(b) the modulation of the density in the middle of the cloud (high density area) moves rapidly
in the $y$-direction due to strong scattering, leading to two density peaks appearing symmetrically about the $x$-axis for each cloud. Due to the trap potential, self-interference occurs when these two
density maxima move back towards one other as seen in Fig. \ref{fringe1}(d). The most distinct feature in this regime
is that the clouds no longer pass cleanly through each other and regain their original shape after their collision. We can see clearly that there are more structures in the clouds in Fig. \ref{fringe1}(d) than in Fig. \ref{fringe1}(c).
The interference fringes will be completely distorted when the second collision begins and this leads to excitations in the form of vortex pairs and soliton rings as shown in Fig. \ref{exc-lv}.

\begin{figure}[]
  \centering
   \includegraphics[scale=0.7,bb=-80 160 779 650, clip]{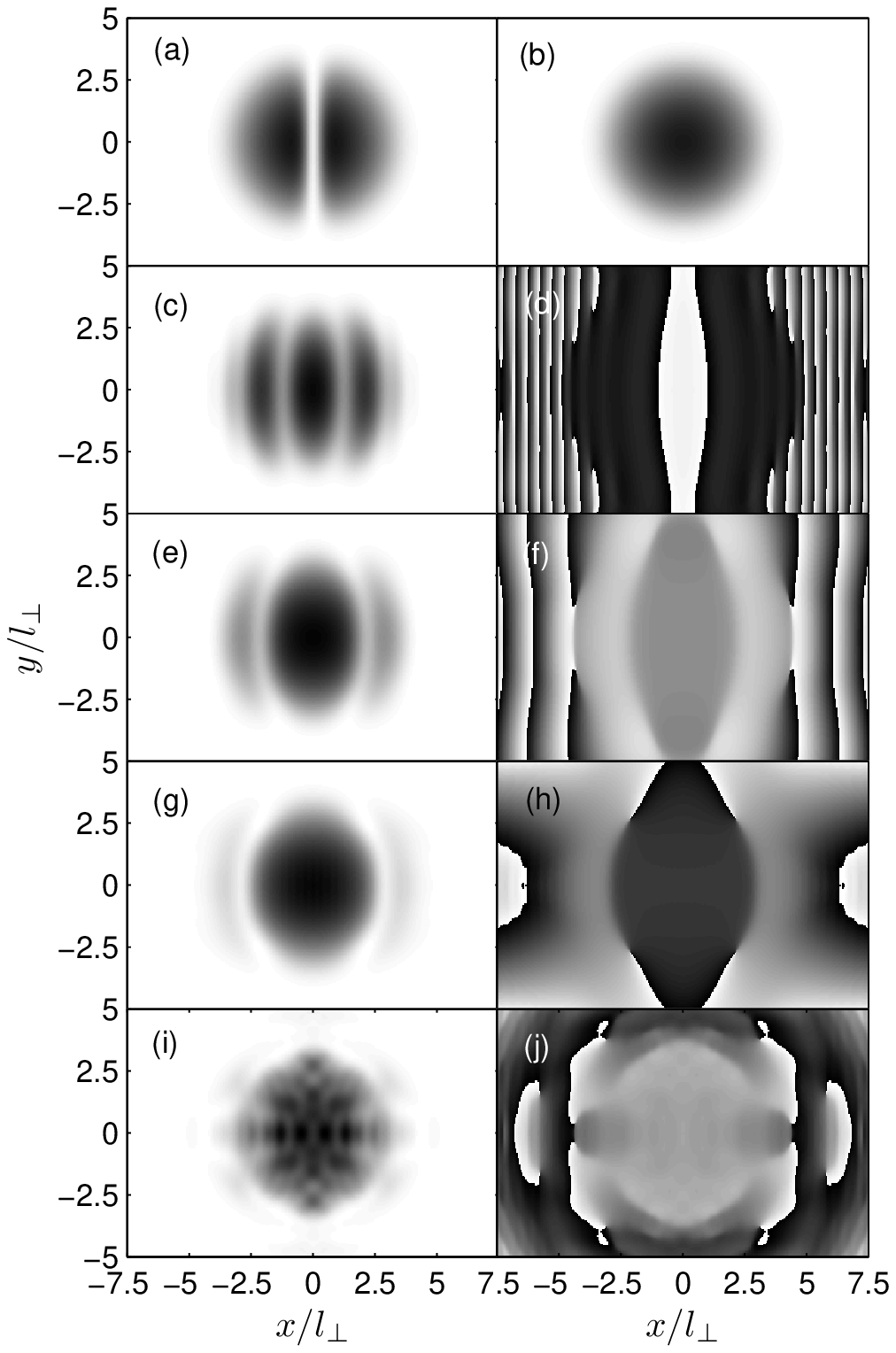}
\caption{Normalized density profiles and phase plots of the condensates prepared in a double well potential which are released into a harmonic potential well at time $t=0$. The initial relative phase is $\theta = 0$. The left panels are the density profiles of the condensates prepared in a double well potential (a) and then released in the harmonic potential (c,~e,~g,~i). (d,~f,~h,~j) are the corresponding phase plots of (c,~e,~g,~i) respectively. (b) is the density profile of the ground state condensate with $N$ atoms in the harmonic potential well. The density profile of the dynamic state (g) is close to the ground state (b), but with some distortion and a few excited atoms around it. Density and phase modulations are obvious in plots (i) and (j). Times are (a) $\tilde t=0$, (c,~d) $\tilde t=0.5$, (e,f) $\tilde t=1.0$, (g,~h) $\tilde t=1.57 $, (i,~j) $\tilde t=10 $. }\label{dw-0-phase}
\end{figure}

In Fig. \ref{collision}, we show quantitatively the scattering effect by plotting the dynamics of $N_y$, which is the number of atoms with $y$-component of momentum $|\tilde{k}_y|>10d\tilde k$ where $2\pi/25$ is the spatial resolution of the simulation. As seen in the figure, the initial value of $\tilde{k_y}$ is 0 because the condensate clouds are set to move along the x-axis. The black solid line is the absolute value of a sinusoidal curve fitted with fixed frequency $\tilde{\omega}=1$ which indicated that there are two collisions during one dipole period. For all three distinct initial conditions there is only one peak during the first half period because the scattering occurs after the two clouds collide at the bottom of the trap ($\tilde t \sim \pi/2$). The peak value increases with decreasing $\Delta$. After the first half period two peaks appear. The first one is induced by the self-interference in the y-direction of each cloud, and the second one arises from the collision of the two clouds. For the underdamped case (the line marked with open circles) the self-interference induced scattering is always weaker than the collision induced scattering before the conditions of the underdamped regime are violated (recall that $\eta_n$ increases with $n$), while it is true for the critically damped case (the line marked with cross and the line marked with dot) only in the second half period. In this intermediate regime the creation of solitons and vortices after the second half period suppress the scattering along the x-direction (collision induced scattering), which makes the second peak lower than the first one.

Generally in the above regimes the initial phase difference of two clouds will not change the physics intrinsically. Especially for the underdamped situation, only the location of the fringes varies with the
relative phase $\theta$, while the period and spacing of the fringes are not changed.
In the underdamped regime, the trajectory is independent of $\theta$ indicating that the initial relative phase will not change the scattering properties at all. In the critically damped regime the only difference between the $\theta=0$ (dotted line) and $\theta=\pi$ trajectories is the peak values. An interesting point is that the peak values of $\theta=\pi$ is always smaller than those of $\theta=0$, which is different from the results for the merging process (overdamped collision) in Ref.\cite{PRL.98.180401} where the atom loss after recombination increased as the relative phase increased from 0 to $\pi$.

If we reduce the distance between two clouds further, the interaction energy will dominate the system. In this overdamped situation the dynamics of the system could be very different from the cases we have considered so far. The two clouds could merge into one condensate after the first collision. In the next section we will look at this nonlinear regime where the initial relative phase plays a more important role.

\begin{figure}[t]
  \centering
   \includegraphics[scale=0.7,bb=-40 250 779 600, clip]{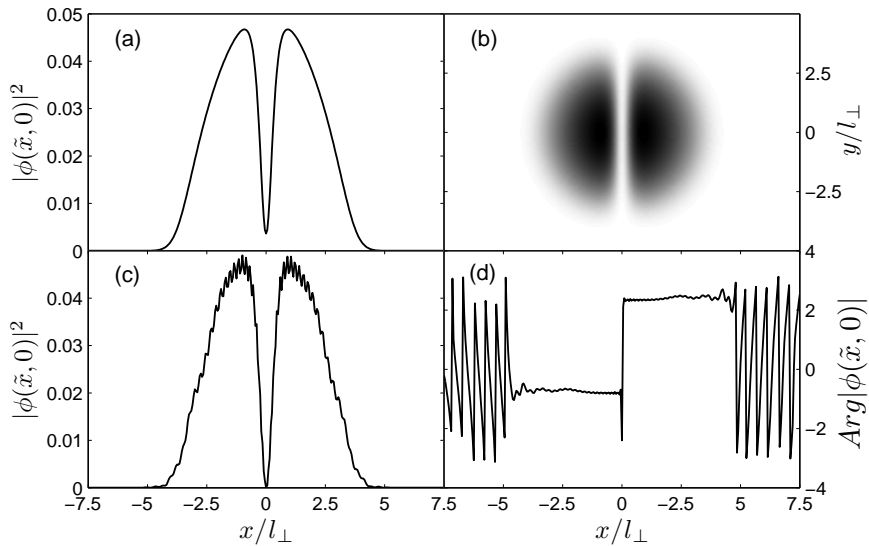}
   \caption{(a) The initial density profile of a condensate prepared in a 2D double well potential at $y = 0$. (b) the top view of the initial density profile.  Density (c) and phase (d) profile at $\tilde{t} =0.5$ and $y = 0$. }\label{dw-pi-initial}
\end{figure}

\subsection{Phase-sensitivity of excitations in overdamped collisions}

In Ref.\cite{PRL.98.180401,PRA.79.063624}, it was reported that for elongated condensates that are tightly confined in the radial
directions, when the central barrier potential is suddenly turned off, two clouds merge into a single cloud. When the initial relative phase between the two clouds is zero, the merged cloud is in its ground state with respect to the trapping potential. When the initial relative phase is $\pi$, the merged condensate contains a dark soliton.

For 2D disk-shaped clouds, the size of the condensate in the merging direction (x-axis in our case) is large. As we have seen, when the separation between the c.m. of the two clouds is comparable to the diameter of one cloud, i.e. $\Delta\sim R_{TF}$, the c.m. kinetic energy that the clouds will have is comparable to the interaction energy. This energy is high enough to prevent the two clouds merging into one condensate after the first collision, as they will not lose sufficient momentum in the merging direction due to scattering. If we want to investigate the merging of 2D condensate clouds we need to enter the overdamped regime where the interaction energy dominates the system, i.e. $\Delta<R_{TF}$. This means that there will be an overlap between the two clouds which depends on the height and width of the central barrier of the trap, $V_{DW}$ (see Eq. (\ref{DW})).

As we have seen, solitons are created during the interference process. We know that soliton stability depends on the nonlinearity and geometry of the medium \cite{Phys.Rep.298.81}. In general, solitons in BECs are always thermodynamically unstable, and dark solitons in BECs are also expected to be inherently dynamically unstable \cite{JPB.33.3983,PRA.62.053606,PRA.60.R2665,PRA.60.3220}. In a uniform system with background density $\rho$, a soliton corresponding to a phase jump of $\theta$ ($0<\theta\leq\pi$) has velocity \cite{soliton,JPB.30.L785}
\begin{equation}\label{velocity}
v_s/v_c = \cos(\theta/2)=(1-\rho_d/\rho)^{1/2},
\end{equation}
where $v_c=(g_{2}\rho/m)^{1/2}$ is the speed of sound and $\rho_d$ is the difference between $\rho$ and the minimum density in the centre of the soliton. For a given density, increasing the relative phase from 0 to $\pi$ will decrease the velocity of the dark soliton. In the limiting case $\theta=\pi$, the velocity of a soliton is $0$.
The formation of vortex pairs inside the combined condensate depends strongly on the velocity of the soliton $v_s$ and the density distribution of condensates.

Firstly, we consider the situation for which the initial relative phase between the condensates is 0.  As shown in Fig. \ref{dw-0-phase}(a), there is an overlap between the cloud in the left well and that in the right well. The dynamics of the two clouds will lead to them interfering and merging after the central barrier is turned off. The spacing of the central fringe increases with time until the two clouds reach their maximum overlap and the shape of the central fringe is close to that of the ground state of a condensate with $N$ atoms in the harmonic trap (see Fig. \ref{dw-0-phase}(g) and (b)). The outer fringes nearly vanish and the two clouds merge into one with some excited atoms being found near the edge of the cloud (see Fig. \ref{dw-0-phase}(g)). The scattering is sufficient, due to the long collision time, to ensure that the combined condensate can not separate any more. The expansion of the condensate is not as obvious as in Fig. \ref{fringe1}(b) due to the small maximum c.m. kinetic energy of the clouds and the fact that the interatomic interaction dominates the system. However the cloud cannot reach its ground state, as would happen in the 1D case, because the trapping frequency in the merging direction is too low. As predicted by Eq. (\ref{velocity}) the dark solitons move at $v_c$ and decay rapidly so the system should be dominated by density wave excitations. In Fig. \ref{dw-0-phase}(i) and (j) we see that a state with density and phase modulations, but without topological defects, is obtained, meaning the system is in a collective excited mode.

\begin{figure}[t]
  \centering
   \includegraphics[scale=0.7,bb=-80 200 779 630, clip]{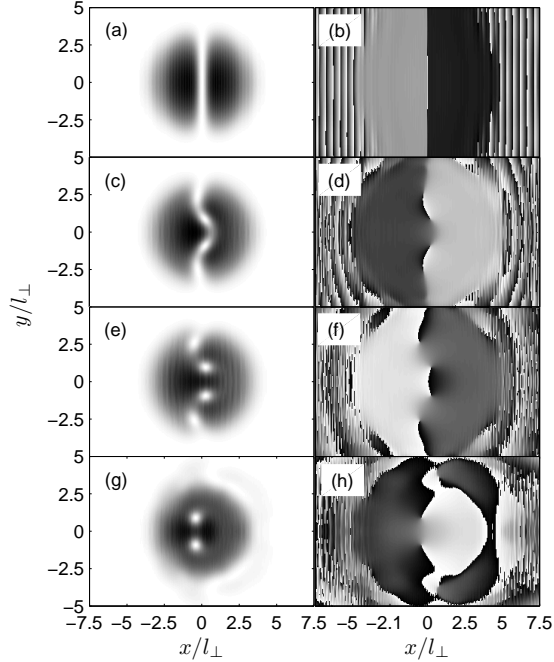}
   \caption{Sequence showing the density profiles (the left panels) and the corresponding phase diagrams (the right panels) for condensates prepared in a double well potential undergoing a sudden merger in a harmonic trap. The initial relative phase is $\theta = \pi$. Times are (a,b) $\tilde t=0.5$, (c,d) $\tilde t=7.5$, (e,f) $\tilde t=8$, (g,h) $\tilde t=15.2 $.}\label{dw-pi-merge}
\end{figure}

Secondly, we consider the situation where the initial relative phase is $\pi$, but with the other conditions kept
unchanged. We find this merging process is completely different to the case where $\theta=0$, and that it is also distinct from the quasi-1D merging of condensates with relative phase $\pi$ reported by Ref.\cite{PRL.98.180401} and Ref.\cite{PRA.79.063624}, where the merging time is too fast (merging occurs in the tight confining direction) to make the soliton decay. In the 2D case, the radial frequency of the trap is weak, and so will not quench mechanical excitations in either $x$- or $y$-direction. As seen in Fig. \ref{dw-pi-initial}(a) and (b), the condensates with relative phase $\pi$ are initially prepared in a double well potential. At time $t=0$, the central gaussian barrier is turned off suddenly. For interference with $\pi$ relative phase there should be two central fringes. Due to the small initial separation of the two clouds, these two central fringes are very fat and nearly keep the original shape of the initial clouds. Our simulations show that two density waves propagate in opposite directions at a reduced amplitude as seen in Fig. \ref{dw-pi-initial}(c). The initial soliton will eventually reach a point where its central density is zero, and the phase difference is $\pi$ between the two parts (see Fig. \ref{dw-pi-initial}(c) and (d)), where a completely dark soliton (black soliton) is created. The velocity of the black soliton is zero. In Fig. \ref{dw-pi-merge}, we give the snapshots of the time evolution of the condensates which show that the black soliton is not as stable as in the merging of elongated condensates. After approximately 143ms ($\tilde{t} \simeq 4.5$) of real-time propagation, the black soliton begins to undulate. It is bent at first, then divides into two from the middle with atoms passing through the centre and propagating from left to right. The atoms in the right cloud are pushed by these additional atoms and then two paths from right to left are formed, one above and one below the centre of the cloud, leading to the decay of the soliton via the snake instability \cite{PRL.76.2262,PRA.54.870}. A pair of vortices form near the centre of the condensate and subsequently the other two vortices nucleate at the edge of the cloud, move around the edge to the left and towards each other, and then annihilate at $y = 0$. We note that due to the initial phase difference of $\pi$ between the two condensates there is a persistent phase jump as shown in Fig. \ref{dw-pi-merge}(f) and (h).

For initial relative phases between 0 and $\pi$, the initial dark soliton propagates in the positive $x$-direction, and curves with the expansion of the central interference fringe. This curvature is similar to that observed in Ref.\cite{Science.287.97}. However the dark soliton will eventually decay into vortex pairs instead of oscillating in the trap.

\begin{figure}[t]
\centering
\includegraphics[scale=0.6,bb=-200 150 779 690, clip]{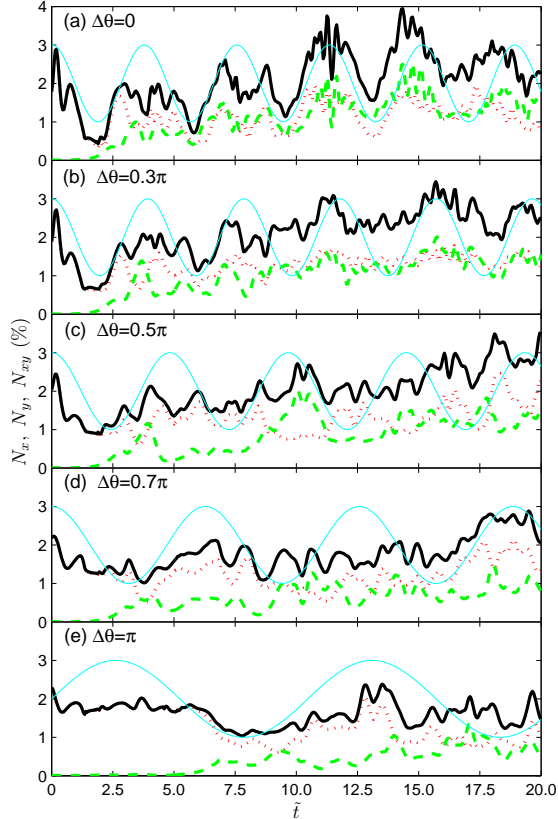}
\caption{(Color online) Oscillations of condensate atom number $N_x$ (red dotted line), $N_y$ (green dashed line) and $N_{xy}$ (thick solid line) for different initial relative phases.  The cyan thin lines are sinusoidal curves fitted with frequencies $\tilde\omega=1.66$ (a), 1.6 (b), 1.3 (c), 1.0 (d), 0.6 (e). }\label{excitation_merge_time}
\end{figure}

The nonuniform density distribution in our system makes the speed of a soliton a function of $y$, which leads to the curvature of the solitons. the formation of vortex pairs inside the combined condensate would require longer lifetimes (i.e. smaller velocity) for the solitons. In Fig. \ref{excitation_merge_time} $N_x$ is the number of atoms with $\left\vert\tilde{k_x}\right\vert>10d\tilde k$, and $N_{xy}$ is the number of atoms with $\left\vert\tilde{k_x}\right\vert$ or $\left\vert\tilde{k_y}\right\vert$ larger than $10d\tilde{k}$. We can see that $N_y$ is almost unchanged until the soliton began to decay (green dashed line), which means that the scattering in the $y-$direction is suppressed by the formation of solitons. The lifetime of solitons increases as $\theta$ is increased from 0 to $\pi$ as predicted by Eq. (\ref{velocity}). When $\theta$ is smaller than $\pi/2$ the oscillation frequencies of $N_x$ and $N_y$ are nearly the same after the decay of solitons as is that of the $N_{xy}$ oscillation. When $\theta = 0$ the velocities of the accompanying density waves are close to $v_c$ \cite{PRL.83.5198} and the corresponding frequencies are close to $2\omega_{\perp}$. As shown in Fig. \ref{excitation_merge_time}(d) and (e), when $\theta$ is above $\pi/2$ the formation of vortices inside the combined condensate as seen in Fig. \ref{dw-pi-merge} (g) will diminish the amplitude of the number oscillation, which means that the presence of vortex pairs in the combined condensate suppresses density waves.

\section{Conclusion}

\


In this paper we have studied, within zero-temperature mean-field theory, the way in which quantum interference fringes can seed the irreversible formation of internal excitations (soliton and vortex textures) in counter-propagating 2D Bose condensed clouds of ultra-cold atoms. Such 2D geometries are of interest for their potential in matter-wave interferometry as they are likely to be less susceptible to thermal and quantum fluctuation effects than their quasi-1D counterparts. Generally, the nature of the dynamical excitations created when two clouds collide or merge depends not only on the initial separation of the clouds but also on their initial relative phase. This is true in both quasi-1D and 2D systems. However the collision and merging processes in the 2D configurations are quite different to those observed in 1D or quasi-1D systems\cite{PRL.98.180401,PRA.79.063624}. We also found that scattering effects are suppressed in 2D, which potentially increases the sensitivity in some interferometric schemes. Furthermore, the relative phase of the atom clouds can be determined over a wide range in the overdamped merging process even at zero temperature. In our 2D calculations we show that in the two limiting cases (underdamped and overdamped regimes) scattering is less effective due to the short collision time or the small c.m. kinetic energy. In the underdamped regime during our simulations the amplitude of $N_{y}$ oscillation is smaller than $4\%$ for the first collision, and is still smaller than $15\%$ even during the third collision. In the overdamped regime $N_{xy}$ is always under $4\%$ as $\tilde t<20$, while the resulting oscillations of the condensate atom number in Ref.\cite{PRL.98.180401} are about $25\%$. This means that an in-trap 2D interferometer can provide better contrast. Only when the system is in the critically damped regime is the scattering sufficient (up to $40\%$ around the third collision), to make the transverse self-interference of each cloud obvious. The combination of the interference patterns in both the $x-$ and $y-$directions renders the dynamical excitations in the system quite complicated.

We have shown that there is no simple linear relation between the formation of topological excitations and the c.m. kinetic energy. The smaller the c.m. kinetic energy, the more important is the initial relative phase. When the kinetic energy dominates, regular, if slightly curved, interference fringes are observed and there are no dynamical excitations after the two clouds are separated. In the critically damped regime vortex formation occurs which is not intrinsically affected by the initial relative phase. In the overdamped regime we observe significant qualitative sensitivity to the initial phase difference. The two clouds can merge into one in this case, while the final states are, however, excited in different ways, depending on the initial relative phase. We found that, only when the initial relative phase is larger than $\pi/2$, will there be a vortex pair which can persist in the combined condensate. Otherwise, all the vortex pairs formed in the merging process will simply move around the edge of the combined condensate and escape so that the system is left with density modulations.

Jo {\it et al} \cite{PRL.98.180401} also showed that the recombination of a split quasi-1D condensate held by an atom chip leads to a heating of the atomic cloud which depends on the relative phase of the two clouds. This qualitatively agrees with the analytical prediction that the recombination process can lead to exponential growth of unstable modes \cite{PRA.68.013611}. Our simulations show that in the underdamped regime the dynamical properties are not sensitive to the relative phase at all. The role of initial phase difference is only limited to the position of the fringes. In the critically damped regime the collective density wave oscillations do not increase with increasing relative phase from 0 to $\pi$. In the overdamped regime the oscillation of $N_{xy}$ is suppressed by the formation of topological excitations (solitons and vortices) inside the clouds when the relative phase is $\pi$.

To conclude, we have identified the ways in which the collision/merger dynamics of a coherent pair of 2D clouds differs from their quasi-1D counterparts. In particular, the way in which the decreased stability of a soliton affects the outcome of collisions at late times. We can expect the mean-field approach that we have used to work well in situations in which thermal and quantum fluctuations are negligible. Of course, this limits the description to low temperatures but also to situations in which the particle density is sufficiently large that the order parameter can be treated as a classical field. Hence, even at zero temperature, quantum fluctuations may be important in the regions close to the minima of the interference fringes and the perimeter of the atom clouds. Thermal fluctuations will suppress the magnitude of the order parameter, reducing the fringe contrast, and cause lateral modulations in the fringe pattern. Thermal effects will also allow the spontaneous nucleation of vortex-antivortex pairs which will proliferate as the Kosterlitz-Thouless transition\cite{kt1,kt2} is approached. In the overdamped regime, this should broaden the sharp change in behaviour seen here at a phase difference of $\pi/2$. Previous theoretical studies of the quantum dynamics of bosonic atoms in a combined harmonic and lattice potential indicated that quantum fluctuations can result in strong dissipation \cite{PRL.93.070401,PRA.72.011601,PRL.95.020402}. We expect, however, that for more moderate and weak interaction strengths, the inclusion of quantum fluctuations will not qualitatively affect the behaviour in the underdamped and, at least at early times, the critically damped regimes but will likely have a strong effect in the overdamped regime where complex textures develop in the merged cloud. Hence the next step in the study of these processes should be the inclusion of such quantum (and thermal) fluctuations using, for example, the finite temperature truncated Wigner approach \cite{twa1, twa2, twa3}.

\begin{acknowledgments}
This work was supported by the EPSRC. We would like to thank Mark Fromhold and Andrew Henning for many useful discussions.
\end{acknowledgments}

\typeout{}

\bibliography{dynamical_excitations_arXiv_revised.bbl}

\end{document}